\documentclass[11pt,reqno]{article}

\usepackage{mathtools}
\usepackage{amsmath}
\usepackage{amsfonts}
\usepackage{amssymb}
\usepackage{amsxtra}
\usepackage{graphicx}
\usepackage{caption}
\usepackage{ushort}
\usepackage{color}
\usepackage{hyperref}
\usepackage[parsep]{collref}
\hypersetup{linktocpage=true}
\usepackage{enumitem}
\usepackage{titlesec}
\usepackage{cite}
\usepackage{float}
 \def\bd{\begin{document}} \def\ed{\end{document}}
\def\ds{\documentstyle} \let\fr=\frac \let\bl=\bigl \let\br=\bigr
\let\Br=\Bigr \let\Bl=\Bigl
\let\bm=\bibitem
\let\na=\nabla
\let\pa=\partial \let\ov=\overline

\newcommand{\be}{\begin{equation}}
\newcommand{\ee}{\end{equation}}
\newcommand{\bea}{\begin{eqnarray}}
\newcommand{\eea}{\end{eqnarray}}
\newcommand{\ba}{\begin{array}}
\newcommand{\ea}{\end{array}}

\def\ft#1#2{{\textstyle{{\scriptstyle #1}\over {\scriptstyle #2}}}}
\def\fft#1#2{{#1 \over #2}}\def\del{\partial}
\def\vp{\varphi}
\def\sst#1{{\scriptscriptstyle #1}}
\def\st#1{{\scriptstyle #1}}
\def\oneone{\rlap 1\mkern4mu{\rm l}}
\def\td{\tilde}
\def\wtd{\widetilde}
\def\ie{{\it i.e.\ }}
\def\iec{{\it i.e.,\ }}
\def\eg{{\it e.g.\ }}
\def\egc{{\it e.g.,\ }}
\def\dalemb#1#2{{\vbox{\hrule height .#2pt
        \hbox{\vrule width.#2pt height#1pt \kern#1pt
                \vrule width.#2pt}
        \hrule height.#2pt}}}
\def\smsquare{\mathord{\dalemb{6.8}{7}\hbox{\hskip1pt}}}
\newcommand{\ho}[1]{$\, ^{#1}$}
\newcommand{\hoch}[1]{$\, ^{#1}$}
\newcommand{\ra}{\rightarrow}
\newcommand{\lra}{\longrightarrow}
\newcommand{\Lra}{\Leftrightarrow}
\newcommand{\ap}{\alpha^\prime}
\newcommand{\bp}{\tilde \beta^\prime}
\newcommand{\tr}{{\rm tr} }
\newcommand{\Tr}{{\rm Tr} }
\newcommand{\supp}{{\rm supp}}
\def\0{{\sst{(0)}}}
\def\1{{\sst{(1)}}}
\def\2{{\sst{(2)}}}
\def\3{{\sst{(3)}}}
\def\4{{\sst{(4)}}}
\def\5{{\sst{(5)}}}
\def\6{{\sst{(6)}}}
\def\7{{\sst{(7)}}}
\def\8{{\sst{(8)}}}
\def\9{{\sst{(9)}}}
\def\ten{{\sst{(10)}}}
\def\n{{\sst{(n)}}}
\def\cA{{{\cal A}}}
\def\cF{{{\cal F}}}
\def\tV{\widetilde V}
\def\tW{\widetilde W}
\def\tH{\widetilde H}
\def\tE{\widetilde E}
\def\tF{\widetilde F}
\def\tA{\widetilde A}
\def\im{{{\rm i}}}
\def\tY{{{\wtd Y}}}
\def\ep{{\epsilon}}
\def\vep{{\varepsilon}}
\def\bD{{{\bar D}}}
\def\alp{{{\a'}^3}}
\def\bD{{{\bar D}}}
\def\R{{{\mathbb R}}}
\def\C{{{\mathbb C}}}
\def\E{{{\mathbb E}}}
\def\H{{{\mathbb H}}}
\def\CP{{{\mathbb C}{\mathbb P}}}
\def\RP{{{\mathbb R}{\mathbb P}}}
\def\Z{{{\mathbb Z}}}
\def\bA{{{\mathbb A}}}
\def\bB{{{\mathbb B}}}
\def\bC{{{\mathbb C}}}
\def\bR{{{\mathbb R}}}
\def\bD{{{\mathbb D}}}
\def\bI{{{\mathbb I}}}
\def\bE{{{\mathbb E}}}
\def\bM{{{\mathbb M}}}
\def\bZ{{{\mathbb Z}}}
\def\Re{{{\frak{Re}}}}
\def\Im{{{\frak{Im}}}}
\def\cosec{{\,\hbox{cosec}\,}}
\def\Gm{{\Gamma_{\!\! -}}}
\def\Gp{{\Gamma_{\!\! +}}}

\def\cosech{{\hbox{cosech}}}
\def\sech{{\hbox{sech}}}
\newcommand{\cB}{\mathcal{B}}
\newcommand{\cD}{\mathcal{D}}
\newcommand{\cE}{\mathcal{E}}
\newcommand{\cH}{\mathcal{H}}
\newcommand{\cJ}{\mathcal{J}}
\newcommand{\cK}{\mathcal{K}}
\newcommand{\cL}{\mathcal{L}}
\newcommand{\cM}{\mathcal{M}}
\newcommand{\cO}{\mathcal{O}}
\newcommand{\cS}{\mathcal{S}}
\newcommand{\cU}{\mathcal{U}}
\newcommand{\cX}{\mathcal{X}}
\newcommand{\fg}{\mathfrak{g}}
\newcommand{\fh}{\mathfrak{h}}
\newcommand{\hg}{\hat{g}}
\newcommand{\hA}{{\hat{A}}}
\newcommand{\hcO}{\hat{\cO}}
\newcommand{\hcU}{\hat{\cU}}
\newcommand{\hPhi}{\hat{\Phi}}
\newcommand{\hPsi}{\hat{\Psi}}
\newcommand{\hLambda}{\hat{\Lambda}}
\newcommand{\oA}{\overline{A}}
\newcommand{\ophi}{\overline{\phi}}
\newcommand{\opsi}{\overline{\psi}}
\newcommand{\olambda}{\overline{\lambda}}
\newcommand{\tpsi}{\tilde{\psi}}

\newcommand{\dbar}{{\mathchar'26\mkern-12mu d}}

\DeclareMathOperator{\U}{U}

\newcommand{\caltech}{\it Walter Burke Institute for Theoretical Physics, California Institute of Technology, Pasadena, CA 91125}

\newcommand{\brandeis}{\it Physics Department, Brandeis University, Waltham, MA 02454}

\newcommand{\imperial}{\it The Blackett Laboratory, Imperial College London\\
Prince Consort Road, London SW7 2AZ}

\newcommand{\auth}{
C. W. Erickson\,\footnote{\,christopher.erickson16@imperial.ac.uk},
Rahim Leung\,\footnote{\,rahim.leung14@imperial.ac.uk},
and K. S. Stelle\,\footnote{\,k.stelle@imperial.ac.uk}} 

\let\oldabstract\abstract
\let\oldendabstract\endabstract
\makeatletter
\renewenvironment{abstract}
{\renewenvironment{quotation}%
               {\list{}{\addtolength{\leftmargin}{1em} 
                        \listparindent 1.5em%
                        \itemindent    \listparindent%
                        \rightmargin   \leftmargin%
                        \parsep        \z@ \@plus\p@}%
                \item\relax}%
               {\endlist}%
\oldabstract}
{\oldendabstract}
\makeatother

\numberwithin{equation}{subsection}

\let\oldsection\section
\renewcommand{\section}{\renewcommand{\theequation}{\thesection.\arabic{equation}}\oldsection}
\let\oldsubsection\subsection
\renewcommand{\subsection}{\renewcommand{\theequation}{\thesubsection.\arabic{equation}}\oldsubsection}

\titlespacing*{\section}
{0pt}{5.5ex plus 1ex minus .2ex}{4.3ex plus .2ex}
\titlespacing*{\subsection}
{0pt}{5.5ex plus 1ex minus .2ex}{4.3ex plus .2ex}

\begin{document}
\setcounter{page}{0}
\thispagestyle{empty}
\begin{flushright}
\hfill{
Imperial/TP/2022/KS/01}\\
\end{flushright} 
\vspace{20pt}

\begin{center}  

{\Large {\bf Higgs Effect Without Lunch}\footnotesize{\hoch\clubsuit}}

\vspace{20pt}

\auth

\vspace{7pt}
\imperial

\end{center}

\vspace{1.5cm}

\begin{abstract}

Reduction in effective spacetime dimensionality can occur in field-theory models more general than the widely studied dimensional reductions based on technically consistent truncations. Situations where wavefunction factors depend nontrivially on coordinates transverse to the effective lower dimension can give rise to unusual patterns of gauge symmetry breaking. Leading-order gauge modes can be left massless, but naturally occurring Stueckelberg modes can couple importantly at quartic order and higher, thus generating a ``covert'' pattern of gauge symmetry breaking. Such a situation is illustrated in a five-dimensional model of scalar electrodynamics in which one spatial dimension is taken to be an interval with Dirichlet/Robin boundary conditions on opposing ends. This simple model illuminates a mechanism which also has been found in gravitational braneworld scenarios.

\end{abstract}
\vfill\leftline{}\vfill
{\footnotesize
\hoch\clubsuit}
Contribution to {\it The Future of Mathematical Cosmology}, Philosophical Transactions A
\pagebreak

\tableofcontents
\addtocontents{toc}{\protect\setcounter{tocdepth}{2}}
\newpage
\pagenumbering{arabic}
\setcounter{page}{1}
\setcounter{footnote}{0}

\section{Vacuum structure and the Higgs Mechanism}

This article is about some of the unusual symmetry-breaking effects that can occur in systems whose low-energy effective dynamics is best described in a lowered spacetime dimensionality.

The phenomenon of symmetry breaking triggered by a non-symmetric vacuum in a theory with a continuous local symmetry is a cornerstone of the Standard Model's unification of the weak and electromagnetic interactions \cite{Englert:1964et,Higgs:1964pj,Guralnik:1964eu,Kibble:1967sv}, as well as of candidate unifying extensions to include the strong interactions (\eg \cite{Georgi:1974sy}). In a now-traditional description of this phenomenon, a feast occurs: the symmetry orbit of vacuum states, which in a rigidly symmetric theory would give rise to Goldstone modes, gives rise in a locally symmetric gauge theory to masses for the gauge fields corresponding to the broken symmetry generators. The broken-generator gauge fields thus ``eat'' the corresponding Goldstone modes, and become heavy.

Symmetry breaking can also occur in the context of a Kaluza-Klein dimensional reduction, either in a gravitational or a non-gravitational model. In a traditional reduction scenario with a compact extra spacetime dimension or set of dimensions, fields are restricted to be independent of the coordinates parametrising the extra dimensions. One speaks of a ``consistent truncation'' in cases where such a restriction may be made fully compatibly with the set of higher-dimensional field equations, so that general solutions to the restricted lower-dimensional theory can be ``lifted'' (or, some say, ``oxidised'') to solutions of the higher-dimensional theory. Such situations are mathematically interesting, and they have been the subject of considerable attention in the field-theory literature. A generalisation of the traditional Kaluza-Klein construction allows for a limited form of dependence on the reduction space  coordinates, while still preserving the truncation consistency -- known as the Scherk-Schwarz mechanism \cite{Scherk:1979zr}. In such a case, fields that always appear differentiated in the original higher-dimensional action can be allowed to have a linear dependence on a reduction coordinate, so their derivatives (or, \eg for gauge fields, field strengths) remain constant over the reduction space.

Despite the seemingly virtuous terminology, ``consistent truncations'' are not physically essential for dimensional reductions. What is more essential is that one be able to make sense of the effective lower-dimensional theory, at least within an appropriate range of length or energy scales. Interactions with higher modes which could simply be set to zero in a consistent truncation can, in a non-consistent case, give rise to small corrections to the leading-order effective theory, \eg of a short-range Yukawa form for massive-field exchanges, or of higher-derivative form but suppressed by inverse powers of higher-mode masses \cite{Duff:1989cr}.

Here, the focus will be on another type of symmetry breaking, in which the vacuum structure does not generate masses for gauge fields in the lower dimension, but in which local symmetry can be broken in a more surreptitious or ``covert'' way -- showing up mainly in the numerical values of interaction-term coefficients at higher orders in a perturbative expansion of the effective theory -- typically at fourth order in the action, and higher \cite{Deser:2019yig}. 

A model where this can be seen \cite{Crampton:2014hia} is in the effective theory for a Type IIA supergravity reduction from ten spacetime dimensions to four, involving the noncompact space ${\cal H}^{(2,2)}$ \cite{Cvetic:2003xr}. Although not yielding a consistent truncation down to $D=4$, an effective four-dimensional theory of supergravity nonetheless emerges thanks to the particular structure of the transverse-space differential equations in a separation-of-variables construction for solutions to the full $D=10$ theory. 

When cast into the form of a Schr\"odinger equation, that transverse problem has a P\"oschl-Teller integrable structure \cite{Poeschl:1933}. 
This Schr\"odinger problem generates a transverse-mode spectrum with a single zero-eigenvalue bound state, then a  gap, and then a continuum of scattering states. That Schr\"odinger-problem eigenvalues translate to mass levels in the lower-dimensional effective theory. It is the existence of the mass gap that allows for a sensible interpretation as a lower-dimensional theory, at least for energies below the scale of the mass gap. The ``vacuum'' solution in this sector of Type IIA supergravity is a lift to $D=10$ of the vacuum solution of a six-dimensional model constructed in 1984 by Salam and Sezgin \cite{Salam:1984cj}, a solution having the spacetime structure of $D=4$ Minkowski spacetime times an $S^2$ two-sphere. An effective $D=4$ supergravity theory then arises (at leading order) for massless fluctuations about this vacuum background, fluctuations incorporating the P\"oschl-Teller zero mode in the transverse part of the wavefunction.

The happy feature of integrability for that transverse Schr\"odinger problem also allows for the explicit calculation of transverse-space integrals giving the coefficients of interaction terms involving the massless modes of the $D=4$ effective theory. Then a funny thing happens: although the $D=4$ effective theory has the massless spectrum of a supergravity theory, beginning in the action at quadratic order with a Fierz-Pauli spin-two Lagrangian, the interaction coefficients at fourth order and higher in fields turn out not to have the values one expects for a generally-covariant theory in an order-by-order field expansion \cite{Arnowitt:1962hi,Weinberg:1965rz}.

This ``funny numbers'' phenomenon is the defining feature of covert symmetry breaking in an effective theory where the lower-dimensional gauge or gravitational fields remain otherwise massless: a Higgs effect without lunch. A full exploration of how this works in the Type IIA model of Refs \cite{Salam:1984cj,Cvetic:2003xr,Crampton:2014hia} becomes quite involved. However, simpler toy models of the phenomenon can be constructed by declaring non-standard boundary conditions for fields in the transverse space dimensions. This was done for a simple illustrative Maxwell-plus-scalar model in Ref.\ \cite{Erickson:2020oda}, which shows the essential features of such a covert symmetry breaking scenario, and to which we next turn.

\section{Massless effective theory in a non-standard reduction:\newline $D=5$ Dirichlet-Robin electrodynamics}\label{sec:effthy}

\subsection{Maxwell action, boundary conditions and gauge symmetries}

Instead of the technically rather involved Type IIA supergravity embedding of the Salam-Sezgin model, one can create an illustrative dimensional reduction model in which fluctuation fields need to have a non-trivial dependence on an extra `transverse' coordinate. Since this will not generate a consistent reduction in the technical sense, the dimensional reduction will need to be understood in the sense of a low-energy effective theory where lower dimensional dynamics is dominant.

Start with Maxwell theory in a five-dimensional spacetime with structure 
\be
    \cM_5 = \bR^{1,3}\times \bI
\ee
in which the intention will be to choose non-standard boundary conditions on fields at the $z=1$ end of the $\bI = [0,\,1]$  interval. The $D=5$ spacetime metric is 
\be
    {ds_5}^2 = \eta_{MN} dX^M dX^N = \eta_{\mu\nu}dx^\mu dx^\nu + dz^2\;,
\ee
denoting the reduction coordinate by $z$ and the $d=4$ reduced spacetime coordinates by $x^\mu$.

The Maxwell gauge fields $A_\mu(x,z), A_z(x,z)$ will satisfy standard $D=5$ equations of motion and the $A_\mu$ components will be assigned standard Dirichlet boundary conditions at the $z=0$ end of the $\bI = [0,\,1]$  interval, but at the $z=1$ end they will be assigned Robin boundary conditions. The Dirichlet/Robin boundary conditions chosen for $A_\mu$ are accordingly
\be 
A_\mu(x,0) = 0 \,,\quad (\partial_z - 1)A_\mu(x,1) = 0 \,. \label{eq:drbcs}
\ee
For the $A_z$ component of the $D=5$ gauge field, it will prove to be necessary for its transverse-space derivative $\partial_z A_z$ to satisfy the boundary conditions \eqref{eq:drbcs}. In the $d=4$ Minkowski subspace directions, all fields and their associated derivatives will be required to fall off as usual at spatial infinity.

The corresponding $D=5$ Maxwell action also needs to include a boundary term at the $z=1$ end:
\begin{align}
S[A_\mu, A_z] &= S_{\text{Max}}[A_\mu,A_z]+S_{\text{BT}}[A_\mu,A_z] \nonumber \\
&= \int d^4x\int_0^1 dz \Big(-\frac{1}{4}F_{\mu\nu}F^{\mu\nu} - \frac{1}{2}F_{\mu z}F^{\mu z}\Big) + \frac{1}{2}\int d^4x\, F_{\mu z}F^{\mu z}\Big\rvert_{z=1} \,, \label{eq:iac}
\end{align}
where $F_{\mu\nu} = \partial_\mu A_\nu - \partial_\nu A_\mu$, and $F_{\mu z}= \partial_\mu A_z - \partial_zA_\mu$. The $D=5$ action \eqref{eq:iac} is invariant under standard $\U(1)$ gauge transformations with gauge parameter $\Lambda(x,z)$\,:
\be
A_\mu \mapsto A_\mu + \partial_\mu\Lambda \,,\quad A_z \mapsto A_z + \partial_z\Lambda \,. \label{U1gaugetransf}
\ee
Varying the action \eqref{eq:iac} while taking care in imposing the transverse boundary conditions \eqref{eq:drbcs} and integrating by parts on the Minkowski boundary at infinity gives rise to a standard set of Maxwell equations
\begin{align}
A_\mu: &\quad \big(\Box_4 + \partial_z^2\big)A_\mu - \partial_\mu\partial^\nu A_\nu - \partial_\mu\partial_z A_z = 0 \,, \label{eq:wveqn} \\
A_z: &\quad \Box_4A_z - \partial_z\partial^\mu A_\mu = 0 \,, \label{eq:zeqn}
\end{align}
where $\Box_4=\partial_\mu\partial^\mu$. Including the boundary term $S_{\text{BT}}$ in the action \eqref{eq:iac} is necessary in order to allow the Robin condition for the field $A_{\mu}$ to be incorporated into a well-posed variational problem generating the field equations (\ref{eq:wveqn},\,\ref{eq:zeqn}).

In order for the boundary conditions \eqref{eq:drbcs} on $A_\mu$ to be gauge invariant, one requires the following restrictions on the form of the local gauge parameter $\Lambda$:
\be
\Lambda(x,0) = c_1 \,,\quad (\partial_z - 1)\Lambda(x,1)=c_2 \,, \label{eq:residualu1}
\ee
\noindent where $c_1$ and $c_2$ are constants which we will generally take to vanish. 

In addition to the local gauge symmetry \eqref{U1gaugetransf}, the Dirichlet/Robin boundary conditions \eqref{eq:drbcs} for $A_\mu$ allow an additional but restricted `harmonic' gauge symmetry
\be
A_\mu \mapsto A_\mu + \partial_\mu \Gamma \,, \quad A_z \mapsto A_z \,, \label{eq:harmonic}
\ee
 where $\Box_4 \Gamma = 0$ and $\partial^2_z \Gamma = 0$. Invariance of the boundary conditions on $A_\mu$ under this harmonic symmetry requires
\be
\Gamma(x,0) = c_3 \,,\quad (\partial_z - 1)\Gamma(x,1)=c_4 \,,
\ee
where generally we will again take the constants $c_3$ and $c_4$ to vanish.

\subsection{Mode expansions}\label{ssec:modexp}

Now let's expand the $D=5$ gauge fields in a mode expansion so that we can derive an effective theory for the leading-order $d=4$ modes.  For $A_\mu$ satisfying the boundary conditions \eqref{eq:drbcs}, one needs an expansion basis satisfying the same boundary conditions. In order to solve Equation \eqref{eq:wveqn} by separation of variables, 
\be
A_\mu(x,z) = \sum_{\ell=0}^\infty a^{(\ell)}_\mu(x)\xi_\ell(z) \label{eq:amuexp}\,,
\ee
one requires a complete set of basis functions $\xi_\ell\,, \ell=0,1,\ldots,\infty,$ satisfying the transverse wavefunction problem
\be
\xi_\ell''(z) = -\omega_\ell^2\xi_\ell(z) \,,\quad \xi_\ell(0) = 0 \,,\quad \xi_\ell'(1) - \xi_\ell(1) = 0 \,
\ee
where $\xi_\ell'(z)=\partial_z\xi_\ell(z)$. The $L_2$ normalised solutions to this transverse problem are
\bea
\xi_0(z) &=&\sqrt{3}z \,, \quad \xi_\ell(z)= \sqrt2\csc(\omega_\ell)\sin(\omega_\ell z) \,,\quad \ell\in\{1,2,\dots\}\\
\tan \omega_\ell&=&\omega_\ell\,,\quad \omega_\ell>0\,.
\eea

When it comes to $A_z$, the situation is somewhat different. No specific boundary conditions are required in deriving \eqref{eq:zeqn} from the variation of the action \eqref{eq:iac} because the only term containing $\delta A_z$ on the boundaries of $\bI$ already vanishes when the equations of motion are satisfied. One learns the behaviour of $A_z$ on the boundaries instead directly from compatibility with Equation \eqref{eq:wveqn}. Substituting the expansion \eqref{eq:amuexp} into \eqref{eq:wveqn}, one obtains
\be
\sum_{\ell=0}^\infty\Big(\big(\Box_d - \omega^2_\ell\big)a_\mu^{(\ell)}-\partial_\mu\partial^\nu a^{(\ell)}_{\nu}\Big)\xi_\ell(z) - \partial_\mu\partial_zA_z = 0 \,,
\ee
where $\omega_0= 0$. Accordingly, one sees that $\partial_z A_z$ must lie within the span of the $\xi_\ell(z)$ basis, \ie
\be
\partial_zA_z(x,z) = \sum_{\ell=0}^\infty b^{(\ell)}(x)\xi_\ell(z)\,.\label{eq:dAexpansion}
\ee
Integrating this equation and noting that for $\ell>0$ the indefinite integral of $\xi_\ell(z)$ is proportional to its derivative, one has the expansion
\be
A_z(x,z) = h(x)\zeta(z)+ \sum_{\ell=0}^\infty g^{(\ell)}(x)\xi'_\ell(z) \,, \label{eq:azexp}
\ee
where it is convenient to choose $\zeta(z) =\frac{3\sqrt3}{10}-\frac{\sqrt{3}}2z^2$ satisfying $\zeta'(z)=-\xi_0(z)$, to complete the expansion of $\partial_zA_z$ in \eqref{eq:dAexpansion}.

\subsection{Leading-order $d=4$ effective field theory: no lunch}\label{sec:effectivefieldthy}

Using the mode expansions (\ref{eq:amuexp},\,\ref{eq:azexp}) in the $D=5$ field equations (\ref{eq:wveqn},\,\ref{eq:zeqn}) one obtains an equivalent formulation of the theory in terms of $d=4$ expansion fields. Our main interest here will be the structure of the leading-order theory. This may be done either by working directly with the $D=5$ action \eqref{eq:iac} and using orthonormality properties of the Section \ref{ssec:modexp} mode expansions to separate the leading-order terms or by making the mode expansions in the $D=5$ field equations (\ref{eq:wveqn},\ref{eq:zeqn}), separating out the leading-order modes and then reconstructing the leading effective action. Either way, one obtains simply
\be
I_{\sst{\rm Eff}}=\int d^4x\left(-\ft14f_{\mu\nu}f^{\mu\nu} + \partial^\mu h(\partial_\mu g - a_\mu)\right)\,,\label{eq:effact}
\ee
where $f_{\mu\nu}=\partial_\mu a_\nu-\partial_\nu a_\mu$ and $a_\mu(x)$ and $g(x)$ are just the $\ell=0$ modes in the expansions \eqref{eq:amuexp} and \eqref{eq:azexp} with the $\ell=0$ index dropped. The higher $\ell\ge1$ modes excluded from the leading-order action \eqref{eq:effact} are all massive, as in an ordinary Kaluza-Klein expansion. Thus, one expects \eqref{eq:effact} to accurately describe the leading gauge-field dynamics of the system at energies well below the mass of the lightest ($\ell=1$) such higher mode, even when interactions with sources are introduced.

What does the action \eqref{eq:effact} describe? A key point to note is that expanding in terms of the $\xi_i(z)$ and $\zeta(z)$ transverse wave-function basis has not produced a leading-order $d=4$ system with a massive vector field. So this reduction has not generated a standard Higgs mechanism -- the leading-order vector field $a_\mu(x)=a_\mu^0(x)$ remains massless. So there is no lunch-feast going on here. 

The next question is: what are the degrees of freedom described by the effective action \eqref{eq:effact}? The answer involves the $\U(1)$ gauge symmetry \eqref{U1gaugetransf} and the harmonic symmetry \eqref{eq:harmonic}. For the leading-order modes, the $\U(1)$ gauge transformations are, expanding $\Lambda(x,z)$ also in the $\xi_i(z)$ basis with $\lambda(x)=\lambda^0(x)$,
\be
\delta a_\mu(x)=\partial_\mu \lambda(x)\,, \quad \delta g(x)=\lambda(x)\,,\quad \delta h(x)=0\,.
\ee
Notice the r\^ole of the $g(x)$ field in \eqref{eq:effact}: it accompanies the otherwise non-gauge-invariant $\partial^\mu h\, a_\mu$ term to make the $\U(1)$ gauge-invariant combination $\partial^\mu h(\partial_\mu g - a_\mu)$. This is a place where reduction with a non-standard transverse wavefunction zero mode has had an effect: the $g(x)$ field is behaving like a Stueckelberg field \cite{Stueckelberg:1938} repairing an otherwise non-gauge-invariant term. Unlike the original Stueckelberg implementation of such a field in a vector field mass term, however, the $g(x)$ field in \eqref{eq:effact} does not participate in the generation of an effective-theory mass.

To analyse the dynamical radiative degrees of freedom of the system, we employ Fourier analysis on the source-free field equations following from \eqref{eq:effact}\,:
\bea
\partial^\mu(\partial_\mu a_\nu-\partial_\nu a_\mu)-\partial_\nu h&=&0\label{eq:aeq}\\
\Box_4 g - \partial^\mu a_\mu&=&0\label{eq:geq}\\
\Box_4 h &=&0\label{eq:heq}\,.
\eea
Fourier transforming the fields
\be
a_\mu(x)=\int\!\! d^4p \exp(-ip_\nu x^\nu)a_\mu(p)\,,\  g(x)=\int\!\! d^4p \exp(-ip_\nu x^\nu)g(p)\,,\  h(x)=\int\!\! d^4p \exp(-ip_\nu x^\nu)h(p)
\ee
gives the momentum-space equations
\bea
p^2a_\nu-p_\nu p^\mu a_\mu-ip_\nu h&=&0\label{eq:amomspfe}\\
p^2 g-ip^\mu a_\mu &=&0\label{eq:rmomspfe}\\
p^2 h&=&0\,.\label{eq:hmomspfe}
\eea
Now decompose $a_\mu(p)=\tilde a_\mu(p)+p_\mu \kappa(p)$ where at every point $p^\mu$ in momentum space the vectors $\tilde a_\mu(p)$ and $p_\mu$ are taken to be linearly independent. Inserting this decomposition into Equations (\ref{eq:amomspfe}, \ref{eq:rmomspfe}) one obtains
\bea
p^2\tilde a_\nu-p_\nu(p^\mu \tilde a_\mu+ih)&=&0\label{eq:tildeapindep}\\
p^2 g-ip^\mu \tilde a_\mu-i p^2\kappa &=&0\label{eq:gakapparel}\,.
\eea
Linear independence of the vectors $\tilde a_\nu(p)$ and $p_\nu$ at the point $p^\mu$ in momentum space in Equations (\ref{eq:tildeapindep},\,\ref{eq:gakapparel},\,\ref{eq:hmomspfe}) then implies 
\bea
p^2\tilde a_\nu&=&0\label{eq:suppatilde}\\
p^\mu\tilde a_\mu+ih&=&0\label{eq:ahrel}\\
p^2g-ip^\mu\tilde a_\mu-ip^2\kappa&=&0\label{eq:gakapparel}\\
p^2h&=&0\label{eq:supph}\,.
\eea

From Equation \eqref{eq:suppatilde}, one learns that the support of $\tilde a_\nu(p)$ is only on the lightcone, \ie
$\supp(\tilde a_\nu(p))=\{p_{\mu}|p^{2}=0\}$. Similarly, from \eqref{eq:supph} one also learns that $\supp(h(p))=\{p_{\mu}|p^{2}=0\}$. Considering Equation \eqref{eq:gakapparel} on the $p^2=0$ lightcone subspace then implies $(p^\mu\tilde a_\mu)|_{p^2=0}
=0$, and consequently $p^\mu\tilde a_\mu(p)=0$ throughout momentum space. However, Equation \eqref{eq:ahrel} then implies $h(p)=0$ except for $p^\mu=0$ (remembering the $p_\nu$ factor in \eqref{eq:tildeapindep}). Consequently, one has it that $h(x)=\hbox{constant}$ in $d=4$ Minkowski space, and is thus non-dynamical.

The remaining dynamical content of the theory resides in Equations (\ref{eq:suppatilde},\,\ref{eq:gakapparel}) after setting $p_\nu h(p)=0$:
\bea
p^2\tilde a_\nu&=&0\label{eq:suppatildered}\\
p^2g -ip^2\kappa&=&0\label{eq:gakapparelred}\,.
\eea
These are still $\U(1)$ gauge invariant. In momentum space, one has 
$\delta a_\mu(p)=-ip_\mu\lambda$ so $\tilde a_\mu(p)$ doesn't transform, but $\kappa$ does, with $\delta\kappa=-i\lambda$. The transformation of the Stueckelberg field $g(p)$ is $\delta g(p)=\lambda(p)$. Consequently, the combination $g-i\kappa$ is gauge invariant and so Equations (\ref{eq:suppatildered},\,\ref{eq:gakapparelred}) are independently $\U(1)$ gauge invariant.

One can fix the $\U(1)$ gauge symmetry by picking any convenient gauge. One common example is Coulomb gauge, $\partial_i a_i(x)=0$, becoming $p_ia_i=0$ in momentum space. Setting $p_0h=0$ in the $\nu=0$ component of \eqref{eq:amomspfe} then implies $p_ip_ia_0=0$, or in position space $\nabla^2 a_0=0$. Requiring a falloff condition at spatial infinity, $a_0\to0$ as $|x^i|\to\infty$, then implies $a_0=0$. Consequently there remain only the standard two dynamical degrees of freedom in the $a_\mu$ Maxwell field. The momentum-space Lorentz condition is also obtained: $p^\mu a_\mu= -p_oa_0+p_ia_i=0$.

The remaining degree of freedom for the system \eqref{eq:effact} is then seen in Equation \eqref{eq:rmomspfe}: $p^2g(p)=0$. Thus, in Coulomb gauge there is an additional candidate scalar degree of freedom residing in the Stueckelberg field. This would be entirely expected in a standard Kaluza-Klein reduction where dependence on the $z$ reduction coordinate is simply suppressed, in which case the $A_z$ component of the $D=5$ Maxwell field becomes a massless scalar in $d=4$. In the present Dirichlet/Robin system, however, the remaining $g$ field has a more tenuous existence. As we have seen in \eqref{eq:harmonic}, there is one more transformation which we have not yet exploited: $D=5$ pure Maxwell theory with Dirichlet/Robin boundary conditions on the transverse interval $\bI$ also has the harmonic symmetry. By combining the harmonic transformation \eqref{eq:harmonic} with a $\U(1)$ gauge transformation \eqref{U1gaugetransf}, one can trade the $\delta a_\mu=\partial \gamma$ harmonic symmetry transformation of $a_\mu$ for a transformation $\delta g=-\gamma$, leaving $a_\mu$ unchanged. The harmonic condition on $\gamma$ corresponds precisely to the support set of $g$, so one can use the reformulated harmonic transformation to send $g\to0$.

One can summarise the degree-of-freedom count for the reduced effective-theory system described by the action \eqref{eq:effact} as follows: there is no massive lunch, and even the remaining scalar $g$ has a kind of `phantom' (but not ghost!) existence -- barely a crumb on the side. Only the two $d=4$ Maxwell degrees of freedom are unambiguously dynamical. We will investigate this structure further by looking at the Hamiltonian, in order to see what impact the presence of $g$ would have on the conserved energy.

\subsection{Effective-theory energy}

From the leading-order effective action \eqref{eq:effact}, one passes to a canonical form of the action in the standard way, defining a canonical momentum  $\pi_a(x)=\frac{\delta I}{\delta \dot{\psi}_a}$ conjugate to each field $\psi_a$ in \eqref{eq:effact}. This yields
\bea
\pi_i&=&\dot a_i-\partial_i a_0\\
\pi_0&=&0\\
\pi_h&=&a_0-\dot g\\
\pi_g&=&-\dot h\label{eq:pig}
\eea
where $\dot \psi=\frac{\partial\psi}{\partial t}$. The corresponding canonical action
\be
I_{\st{\rm canonical}}=\int dtd^3x(\pi_i\dot a_i+\pi_g\dot g+\pi_h\dot h - {\cal H})\label{eq:canaction}
\ee
has the Hamiltonian density
\be
{\cal H}=\ft12\pi_i\pi_i+\ft14f_{ij}f_{ij}-a_0\partial_i\pi_i+\pi_g(a_0-\pi_h)+\partial_ih(a_i-\partial_ig)\,.\label{eq:hamiltonian}
\ee
One may verify that Hamilton's equations following from variation of the canonical action \eqref{eq:canaction} are a first-order system that is fully equivalent to the Lagrangian field equations (\ref{eq:aeq},\,\ref{eq:geq},\,\ref{eq:heq}).

The system's total energy is the value of $E=\int d^3x{\cal H}$ for solutions to the field equations, as discussed above in Section \ref{sec:effectivefieldthy}. From the Coulomb-gauge analysis there, one has $h=\hbox{constant}$, so from \eqref{eq:pig} one also has $\pi_g=0$. Moreover, the field $a_0$ has no conjugate momentum and its variation yields the constraint $\partial_i\pi_i=0$ just as in standard $d=4$ Maxwell theory. Consequently, for solutions to the effective-theory field equations arising from the action \eqref{eq:effact}, one has the conserved total energy
\be
E=\int d^3x(\ft12\pi_i\pi_i+\ft14f_{ij}f_{ij})\,,\label{eq:totenergy}
\ee
which is simply equal to the standard, positive semidefinite, energy of Maxwell theory alone. Although allowed by the field equations, the $g$ field does not contribute to the total energy. Clearly the fact that $g$ may be removed by the residual harmonic symmetry is the reason for its non-appearance in the total energy \eqref{eq:totenergy}. At least in the source-free theory we have considered so far, the $g$ field may be considered redundant.

\section{Covert symmetry breaking}

\subsection{Scalar electrodynamics}

In Section \ref{sec:effthy} above, we have seen that the unambiguous leading-order dynamics of the $D=5$ system \eqref{eq:iac} is just $d=4$ Maxwell theory. The non-standard Dirichlet/Robin boundary conditions on the interval $\bI$ do give rise to the $h$ and $g$ fields in the leading-order effective action \eqref{eq:effact}, but the analysis of Section \ref{sec:effthy} shows that $h$ has no dynamics while $g$ may be considered redundant, giving no contribution to the total energy. Following Reference \cite{Erickson:2020oda}, we now extend the model by $\U(1)$ gauge-covariantly coupling the system \eqref{eq:iac} also to a $D=5$ complex scalar field $\Phi$:
\begin{align}
S[A_\mu, A_z,\Phi,\overline\Phi] =&\; S_{\text{SQED}}[A_\mu,A_z,\Phi,\overline\Phi] + S_{BT}[A_\mu,A_z] \nonumber \\
=&\int d^dx\int_0^1dz\,\Big(-\frac{1}{4}F_{\mu\nu}F^{\mu\nu}-\frac{1}{2}F_{\mu z}F^{\mu z} - \overline{\big(D_\mu\Phi\big)}D^\mu\Phi\Big) \nonumber \\
&+ \frac{1}{2}\int d^dx\, F_{\mu z}F^{\mu z}\Big\rvert_{z=1} \,, \label{eq:qedac}
\end{align}
where $D_M\Phi = \partial_M\Phi - ie A_M\Phi$, $M=\mu,z$, and where $e$ is the charge of the complex scalar field $\Phi$. For simplicity, one may take the $D=5$ scalar field $\Phi$ to satisfy Dirichlet/Dirichlet boundary conditions on the interval $\bI$.

One may expand $\Phi$ in a transverse basis appropriate to the Dirichlet/Dirichlet boundary conditions, 
\be
\Phi(x,z)= \sum_{n=1}^\infty \phi^{(n)}(x)\theta_n(z) \,,
\ee
where $\{\theta_n(z)=\sqrt2\sin(m_nz)\}$ with $n=1,2,\ldots$ and $m_n=n\pi$.

Under the $\U(1)$ gauge symmetry \eqref{U1gaugetransf}, the $\phi^{(n)}$ complex scalar modes transform in a way that mixes the various $n$ levels:
\be
\phi^{(n)} \mapsto \sum_{m=1}^\infty\exp\big(ie\lambda^{(\ell)}I_\ell\big)^{nm}\phi^{(m)} \,, \label{eq:scalaru1}
\ee
 where the matrix $(I_i)^{nm}$ is 
\be
(I_\ell)^{nm} = I_\ell^{nm} =  \int_0^1dz\,\xi_{\ell}(z)\theta_n(z)\theta_m(z) \,.\label{eq:Iinm}
\ee

One may expand the system \eqref{eq:qedac} into $d=4$ modes either by substituting the expansions of $A_\mu$, $A_z$, and $\Phi$ into the higher-dimensional equations of motion or by inserting these expansions into the higher-dimensional action in order to obtain an action for the $d=4$ formulation of the system. The two procedures give equivalent results. Keeping all expansion modes for $A_\mu$, $A_z$, and $\Phi$, the $\U(1)$ gauge symmetry is maintained to all orders, albeit in a rather complicated way.

Now consider just the leading-order effective theory, keeping just the $\ell=0$ massless gauge sector as described in Section \ref{sec:effthy} and the lightest $n=1$ complex scalar mode. The main point of this illustrative $D=5$ scalar electrodynamics model now comes into focus: it is simple enough that the various mode integrals governing expansions beyond the free theory (\ie for action terms cubic and higher in fields) can straightforwardly be done. In addition to the matrix $I_\ell^{nm}$ given in \eqref{eq:Iinm} one needs also 
\be
I_{k\ell}^{nm} = \int_0^1dz\,\xi_{k}(z)\xi_{\ell}(z)\theta_n(z)\theta_m(z)\,.
\ee
Noting that terms $\Phi\partial_\mu\overline\Phi$ and $\overline\Phi\Phi A_\mu$ obey Dirichlet/Robin conditions, and thus can be expanded in the $\{\xi_\ell(z)\}$ basis, one needs the product rules
\be
\theta_n(z)\theta_m(z) = I_\ell^{nm}\xi_\ell(z) \,,\quad \theta_n(z)\theta_m(z)\xi_k(z)=I_{k\ell}^{nm}\xi_\ell(z) \,. \label{eq:expansionthetasxis}
\ee
The covariant derivative operator for the scalar field is then
\be
D^{nm}_\mu = \delta_{nm}\partial_\mu - ieI_{\ell}^{nm}a^{(\ell)}_\mu \,. \label{eq:covderiv}
\ee

Keeping just the leading $\ell=0$ gauge modes and the $n=1$ modes of the scalar, writing simply $\phi=\phi^1$ and keeping terms up to cubic order in the interacting theory, one has
\be
I_{\sst{\rm Eff}}=\int d^4x\left(-\ft14f_{\mu\nu}f^{\mu\nu} + \partial^\mu h(\partial_\mu g - a_\mu)-\partial_\mu\overline\phi\partial^\mu\phi-\pi^2\overline\phi\phi+ie I_0^{11} a^\mu\left(\overline\phi \partial_\mu \phi-\partial_\mu \overline\phi \phi\right)\right)\;,\label{eq:cubiceffthy}
\ee
where $I_0^{11}=\frac{\sqrt3}2$. One can accordingly identify the $d=4$ effective-theory charge for the leading gauge-scalar coupling to be $q_{\sst{\rm Eff}}=\frac{\sqrt3}2e$ in terms of the charge $e$ of the $D=5$ theory.

\subsection{An unanticipated seagull coefficient}

Continuing on to quartic order in fields, one encounters a key peculiarity of this scalar electrodynamics construction with nonstandard boundary conditions. As always in a model with a $\U(1)$ gauge field coupled to a complex scalar field, one expects to have $\int d^4x a_\mu a^\mu\overline\phi\phi$ `seagull' terms occurring at quartic order. From the effective-theory charge $q_{\sst{\rm Eff}}$ identified at cubic order, one would expect the gauge-field-coupled scalar kinetic term to become $-\int d^4x\overline{\big(D_\mu \phi\big)} D^\mu \phi$ where 
\be
D_\mu = \partial_\mu - iq_{\sst{Eff}}a_\mu\label{eq:usualcovder}
\ee
is the usual gauge covariant derivative. Were that the case, the quartic-order $\int d^4x a_\mu a^\mu\overline\phi\phi$ seagull term coefficient would be just $q_{\sst{\rm Eff}}^2$. That is not what arises from the transverse-space integrals, however. Instead, what one finds involves $I_{00}^{11}=1-\frac3{2\pi^2}\ne (I_{0}^{11})^2 $. This is the phenomenon of covert symmetry breaking: no mass generation occurs as in a standard Higgs mechanism, but symmetry breaking in higher couplings, starting at quartic order in fields, does. 

Of course, if one keeps all gauge and scalar modes, the full $D=5$ theory is retained and no symmetry breaking occurs. But this $D=5$ $\U(1)$ symmetry mixes the various gauge and scalar field mode levels in a complicated way. Covert symmetry breaking is intrinsically a phenomenon related to a low-energy approximation, in which the massive higher modes are not independently excited, but are integrated out using the leading-order results of their field equations.

That such a phenomenon can occur starting at quartic order may be expected on general grounds, as well as the fact that when it does occur, it cannot be fully remedied by field redefinitions \cite{Deser:2019yig}. What can be altered by field redefinitions, however is the presentation of the phenomenon. The leading-order effective theory is obtained by integrating out all higher massive modes. For the gauge-field sector, that process is unambiguous when one restricts attention to the leading $\ell=0$ modes. For the complex scalar, however, some rearrangement is possible. Integrating out massive modes at leading order basically involves discarding derivatives of such a mode in its field equation, but keeping the mass term in what then becomes an algebraic equation at leading order, solving for the mode in question. Field redefinitions involving the Stueckelberg modes $g^{(\ell)}$ change the way the $D=5$ $\U(1)$ symmetry acts on the various modes. If one defines \cite{Erickson:2020oda}
\be
\varphi^{(n)} = \exp(ieg^{(\ell)}I_\ell^{nn})\exp(-ie g^{(k)}I_k)^{nm}\phi^{(m)} \,,\label{eq:fredefs}
\ee
then instead of the $D=5$ $\U(1)$ transformations mixing between $n$ modes of the scalar field, the $\varphi^{(n)}$ transform diagonally and canonically,
\be
\varphi^{(n)}\mapsto\exp(ie\lambda^{(\ell)}I_\ell^{nn})\varphi^{(n)} \,,\label{eq:U1transfsmassive}
\ee
noting that $\exp(ie\lambda^{(\ell)}I_\ell^{nn})$ is a phase and not a matrix. 

Integrating out the $n>1$ massive modes and restricting to the $\ell=0$ gauge modes and the $\varphi=\varphi^{(1)}$ scalar field, the result through cubic terms remains the same as in \eqref{eq:cubiceffthy}, but when keeping terms up to quartic order one has
\be
\begin{split}
\tilde I_{\sst{\rm Eff}}=\int d^4x\Big(&-\ft14 f_{\mu\nu}f^{\mu \nu} -\overline{\big(D_\mu \varphi\big)} D^\mu \varphi - \pi^2 \overline \varphi \varphi+\partial^\mu h\big(\partial_\mu g-a_\mu\big) \\
&- q_{\sst{\rm Eff}}^2\widetilde{I}\left(a_\mu-\partial_\mu g\right)\left(a^\mu-\partial^\mu g\right)\overline \varphi \varphi - q_{\sst{\rm Eff}}^2 \widetilde{X} h^2 \overline \varphi \varphi \Big) \,,\label{eq:caneffaction}
\end{split}
\ee
where $D_\mu$ is the usual covariant derivative \eqref{eq:usualcovder} and
\be
\tilde I =(I^{11}_0)^{-2} \left(I_{00}^{11}-(I_{0}^{11})^2\right) =\ft13-\tfrac{2}{\pi ^2}\,,
\ee
while $\widetilde{X}$ is a more complicated but explicitly calculable positive coefficient. In the effective-action form \eqref{eq:caneffaction}, one sees both the preservation of the $\U(1)$ symmetry by Stueckelberg fields but also its breaking. The top line of \eqref{eq:caneffaction} has the normal $d=4$ gauge-covariant coupling of $a_\mu$ to the complex scalar field $\varphi$, while the second line shows the difference seagull vertex $\left(a_\mu-\partial_\mu g\right)\left(a^\mu-\partial^\mu g\right)\overline \varphi \varphi$ with a coefficient  involving the $\tilde I $ combination which gives the deviation from the standard seagull coefficient. Since this structure involves the $\U(1)$ gauge invariant combination $\left(a^\mu-\partial^\mu g\right)$, it makes the preservation of the $D=5$ $\U(1)$ symmetry manifest, but since this is only achieved via the presence of the Stueckelberg field $g$, it also makes the covert symmetry breaking phenomenon manifest. 

The alternative between the original $\phi^{(n)}$ expansion and the transformed $\varphi^{(n)}$ expansion may be likened to the alternative between the gauge basis and the mass basis in the CKM mechanism \cite{Cabibbo:1963yz,Kobayashi:1973fv} in the Standard Model. In the $\phi^{(n)}$ basis, there are no terms directly mixing the $a_\mu^{(\ell)}$ gauge modes and the $g^{(\ell)}$ Stueckelberg modes, but the transformations of the $\phi^{(n)}$ scalar modes mix between different $n$ levels. In the $\varphi^{(n)}$ basis, on the other hand, the scalar modes transform diagonally and canonically but then there are terms involving products of the $a_\mu^{(\ell)}$ and the $g^{(\ell)}$ Stueckelberg modes together with scalar modes, as one sees in \eqref{eq:caneffaction}.

\section{Outlook}

The key issues that we have considered in this article concern the low-energy effective-theory dynamics of a higher-dimensional theory which has a natural effective-theory interpretation in a lower dimension, but in which transverse wavefunction modes have a nontrivial dependence on a transverse coordinate. Such effective reductions can occur in a wide variety of situations, including the Type IIA supergravity lift of the Salam-Sezgin model as considered in Reference \cite{Crampton:2014hia} as well as in the simpler Dirichlet/Robin scalar electrodynamics considered here and in Reference \cite{Erickson:2020oda}. A take-home message is that there are interesting wider varieties of symmetry breaking mechanisms than the standard BEHGHK mechanism. Moreover, reductions in effective spacetime dimensionality can occur without requiring technically consistent dimensional reductions.

The $D=5$ Dirichlet/Robin scalar electrodynamics model illustrated here has leading-order field dynamics in which massless Maxwell theory is unambiguously part of the  $d=4$ effective theory, but there is also the $g$ scalar mode, which, however, makes no contribution to the conserved energy at leading order and may be considered redundant at that level. The full story of that mode in the interacting theory remains to be clarified, as well as the r\^oles of analogous modes in the IIA supergravity / Salam-Sezgin gravitational model. Another question going beyond what has been considered here is what happens when the Dirichlet/Robin scalar electrodynamics system generates a localised source in five dimensions for the gauge sector. One may expect a transition between near-field $D=5$ behaviour near the source and far-field $d=4$ behaviour away from it. Such a near-field / far-field transition is found when a delta-function source localised in the higher dimension is coupled to the Type IIA / Salam-Sezgin model \cite{Erickson:2021psj}; one might expect similar behaviour in the $D=5$ scalar electrodynamics case.

Models that come close to gauge-symmetric models but which display gauge-symmetry breaking without generating masses for the leading-order gauge modes can have subleading corrections which could be of interest from a variety of different perspectives. One such might be whether these corrections can serve as hints of hitherto unrecognised higher-dimensional spacetime structure. 

\section*{Acknowledgments}

We are grateful to Carl Bender, Stanley Deser, Jonathan Halliwell, Alexander Harrold, Jean-Luc Lehners and Massimo Porrati  for helpful discussions. The work of KSS was supported in part by the STFC under Consolidated Grants ST/P000762/1 and ST/T000791/1 and the work of CWE was supported by the United States Department of Veterans Affairs under the Post 9/11 GI Bill.

\end{document}